\documentclass[epj,twocolumn]{webofc}
\usepackage{amssymb,amsmath}
\usepackage[varg]{txfonts}   % Web of Conferences font
\woctitle{NSRT15}
\newcommand{\beq}{\begin{equation}}
\newcommand{\eeq}{\end{equation}}
\newcommand{\bea}{\begin{eqnarray}}
\newcommand{\eea}{\end{eqnarray}}

\newcommand{\mkrm}[1]{}           % use this to hide what has been removed

\begin{document}

\title{Charge-exchange resonances and restoration of the Wigner SU(4)-symmetry in heavy and superheavy nuclei}

\author {Yu.\ S.\ Lutostansky   \thanks{Lutostansky@yandex.ru}, V.\ N.\ Tikhonov}

\institute{ National Research Centre ``Kurchatov Institute'', 123182, Moscow, Russia}

\abstract { Energies
of the giant Gamow-Teller and analog resonances
- $E_{\rm G}$ and $E_{\rm A}$, are presented,
calculated using the microscopic theory of finite Fermi system. The calculated differences
$\Delta E_{\rm G-A}=E_{\rm G}-E_{\rm A}$  go to zero in heavier
nuclei indicating the restoration of Wigner SU(4)-symmetry. The calculated $\Delta E_{\rm G-A}$ values are in good agreement with the experimental data. The average deviation is 0.30 MeV for the 33 considered nuclei for which experimental data is available. The
$\Delta E_{\rm G-A}$ values were calculated for heavy and superheavy nuclei up to the mass number
 $A = 290$. Using the experimental data for the analog resonances energies, the isotopic dependence
  of the difference of the Coulomb energies of neighboring nuclei isobars analyzed within the
   SU(4)-approach for more than 400 nuclei in the  mass number range of $A$ = 3 - 244. The Wigner
   SU(4)-symmetry restoration for heavy and superheavy nuclei is confirmed. It is shown that
    the restoration of SU(4)-symmetry does not contradict the possibility of the existence of
     the "island of stability" in the region of superheavy nuclei.}

\maketitle

\section{Introduction}
 The possible existence of spin-isospin resonance was first discussed in the works of Ikeda, Fujii, and Fujita in the
  middle of the 1960s as an attempt to explain the observed suppression effect of Gamow-Teller (GT) nuclei - transitions \cite{1,2,3}.
 They called this state the Gamow-Teller resonance (GTR), in analogy with analog resonance (AR) or Fermi resonance.

The energies and probabilities of exciting these resonances were calculated later for a large group
of spherical nuclei \cite{4,5,6} within the theory of finite Fermi systems (TFFS) \cite{7}. The first results
 were published in 1972 \cite{4}, several years before the experimental observation of the GTR. It was
  assumed at first that as a spin-flip state, the GTR should be located above the AR by the average
   energy of spin-orbit splitting $E_{ls}$, and that its width should be on the order of $E_{ls}$.
    It was found later from our calculations that the energy splitting between the GTR and AR was
   not equal to $E_{ls}$, but it decreased for heavy isotopes. The effect of decreasing the gap
    between the GTR and AR energies was first noted in 1973 as a result of calculations for more
   than 70 isotopes \cite{5,6}. We concluded that the Wigner SU(4)-supersymmetry \cite{8} must be restored
   in heavy nuclei
    because GTR and AR belongs to the same supermultiplet in this approach, and their main parameters are the same. Since at the time of GTR prediction its energies have not been measured for different nuclei, then at that time it was impossible to solve the question of experimental verification of the hypothesis of  Wigner SU(4)-symmetry restoration in heavy nuclei.

	At present time there are three methods to test this hypothesis by comparing the calculated
and experimental nuclear data. The first - from the analysis of the degeneration of the
Gamow-Teller and analog (AP) resonances, as in this case, both of the resonance must belong to the same supermultiplet according to SU(4)-approach. The second possibility is related to the realization
 for nuclear masses the Franzini-Radicatti relation   \cite{9}  following from the SU(4)-theory. The third
  one - associated with the analysis of the Coulomb energies of nuclei and their isotopic dependence
\cite{10}. Analysis of these three possibilities and restoration of Wigner supersymmetry was
 considered recently in \cite{11}.

In this paper compares the results of our calculations of the energy difference
 $\Delta E_{\rm G-A}$  between
 GTR - $E_{\rm G}$  and AR - $E_{\rm A}$ with experimental data and investigated
  restoration of Wigner SU(4)-symmetry up to superheavy nuclei with $A$ = 290. The isotopic
  dependence of the difference
 of the Coulomb energies of neighboring nuclei isobars also analyzed for more than 400
 nuclei with known experimental data in the range of mass numbers $A = 3 - 244$.
In connection with the restoration of Wigner supersymmetry in heavy nuclei, it becomes
uncertain interpretation of the spin-orbit splitting and the associated shell structure and
 therefore, the possibility of the existence  of the "island of stability" in the region of superheavy nuclei \cite{12}.
 This problem is also discussed in this paper.

%%%%%%%%%%%%%%%%%%%%%%%%%%%%%%%%%%%%%%%%%%%%%%

\section{Method of calculation }
The Gamow-Teller resonance and other charge-exchange excitations of nuclei are described in the TFFS
 with a system of equations for the effective field \cite{7}. For the GT effective nuclear field, we
  obtain a system of equations that, in the $\lambda$-representation has the form:

%%%%%%%%%%%%%%%%%%%%%%%%%%%%%%%%%%%%%%%%%%%%%%%%%
$$
V_{\lambda \lambda^\prime}=V_{\lambda {\lambda}^\prime}^{\omega}+
\sum_{\lambda_1 \lambda_2} \Gamma^\omega_{\lambda \lambda ^\prime \lambda_1
  \lambda_2} A_{\lambda_1 \lambda_2} V_{\lambda_2  \lambda_1}+
\sum_{\nu_1 \nu_2} \Gamma^\omega_{\lambda \lambda^\prime \nu_1  \nu_2}
 A_{\nu_1 \nu_2} V_{\nu_2  \nu_1} ,$$
$$V_{\nu \nu^\prime}=\sum_{\lambda_1 \lambda_2}
\Gamma^\omega_{\nu \nu^\prime \lambda_1  \lambda_2}
A_{\lambda_1 \lambda_2} V_{\lambda_2  \lambda_1}+
\sum_{\nu_1 \nu_2} \Gamma^\omega_{\nu \nu^\prime \nu_1
  \nu_2} A_{\nu_1 \nu_2} V_{\nu_2  \nu_1},
  $$
$$
A_{\lambda \lambda^\prime}^{(p\overline{n} )}=
\frac{n_\lambda^{\overline{n}} (1-n_{\lambda^\prime}^p)}
{\varepsilon_\lambda^{\overline{n}}-
\varepsilon_{\lambda^\prime}^p+
\omega},\quad
A_{\lambda \lambda^\prime}^{(n\overline{p})}=
\frac{n_\lambda^{\overline{p}}
(1-n_{\lambda^\prime}^n)}
{\varepsilon_\lambda^{\overline{p}}-
\varepsilon_{\lambda^\prime}^n-\omega},
 $$
\begin{eqnarray} \quad V_{\rm GT}^\omega =e_q  {\boldsymbol{\sigma \tau_+}}
\label{e1}
\end{eqnarray}
%%%%%%%%%%%%%%%%%%%%%%%%%%%%%%%%%%
where $n_\lambda$ and $\varepsilon_\lambda$ are, respectively, the occupation numbers
and energies of
$\lambda$-states.
The subscripts $\nu$ are used for the $l$-forbidden part of the interaction.
The system of secular of equations (1) for the charge-exchange excitations of nuclei is
 obtained from the more general \cite{13}, including the effective changes of pairing gap
$\Delta$ in the $ph$- and $pp$-channels. In our case we use the condition
$d^{(1)}_{pn}=d^{(2)}_{pn}=0$ assuming that the effects of changes of the pairing gap in
the external field are negligible, which is justified in this case for the external
fields with zero diagonal elements of \cite{7}. Parameters of single-particle states and
their wave functions are calculated in the shell model for neutrons and protons
separately.  Pairing is taken into account in the single-particle structure by the
replacement $\varepsilon_{\lambda} \rightarrow
E_{\lambda}=\sqrt{\varepsilon^{2}_{\lambda}+\Delta^{2}_{\lambda}}$ as in \cite{14},
where the energy $\Delta_{\lambda}$ is calculated separately for neutrons and
protons. The quasiparticle effective charge $e_{q}$ is $e_{q}=1$ for Fermi (${\boldsymbol{\tau \tau}}$)
 transitions and $e_{q}<1$  for Gamow-Teller $({\boldsymbol{\sigma \tau}})$ transitions. The energies
 of charge-exchange excitations are defined as the eigenvalues $\omega_{i}$ of secular
 equations (1), with the most the collective energy ($\omega_{\rm GTR}>\omega_{i}$), and the
 maximum matrix element $M^{2}_{\rm GTR}\approx e^{2}_{q}3(N-Z)$.

In the calculations we use a local
nucleon-nucleon $\delta$-interaction $\Gamma^\omega$ in the Landau-Migdal form with the
 coupling constants $f_L^\prime$  and  $g_L^\prime$ of the
isospin-isospin (${\boldsymbol{\tau \tau}}$) and the spin-isospin (${\boldsymbol{\sigma \tau}}$) quasi-particle interaction
with $L = 0$.
 For the ($\boldsymbol{\tau\tau}$) coupling constant   the value $f_0^\prime = 1.35$ was used, taken
 from comparison of calculating
 energy splitting between the analog and anti-analog isobaric states (IS) with the
 experimental data for the
  large number of nuclei \cite{15}. For the $({\boldsymbol{\sigma\tau}})$ coupling constant  $g_0^\prime$ in the previous
   calculations \cite{16} we used value $g_0^\prime = 1.22$
 obtained by comparing the difference between the GTR $(E_{\rm G1})$ and IS $(E_{\rm G2})$ energies with the experimental
 data for nine Sb isotopes \cite{17}. Considering that the absolute values of the constants can vary in different
 approaches, the obtained ratio $g_0^\prime /f_0^\prime=0.90\pm 0.03$ is model independent in the TFFS.

The energies of GTR and AR were calculated also in the self-consistent TFFS (used a simplified
 version of \cite{18} with the local interaction
and $m^{\ast}=m$), and in its approximate model version \cite{19} in which solutions were
 obtained in analytical form for collective IS.
For this purpose, we neglect the $l$-forbidden terms in (\ref{e1}) and assume
the constant effective field when collective modes excited.
 For the energy differences
  $\Delta E_{\rm G-A}=E_{\rm G}-E_{\rm A}$
  the solution with $V(r)=const$, normalized on the energy $E_{ls}$, has a form for the
 nuclei with $\Delta E>E_{ls}$ $(x=\Delta E/E_{ls} >1)$:
%%%%%%%%%%%%%%%%%%%%%%%%%%%%%%%%%%%%%
\begin{equation}\label{e2}
 y_0 = \frac{\Delta E_{{\rm G}\mbox{-}{\rm A}}}{E_{ls}}
 \approx  (g_{0}^{'} - f_{0}^{'})x
 + b\frac{1+bg_{0}^{'}}{g_{0}^{'} x(1+c_{\rm A}/x^2 )}
\end{equation}
where  $\Delta E =(4/3) \varepsilon _F (N-Z)/A$ ($\varepsilon _F\approx 40$ MeV).

$$
E_{ls}=\sum_{\lambda _1 \lambda _2} n_{\lambda _1}
 (1-n_{\lambda_{2}})\varepsilon_{\lambda _1 \lambda _2}^{ls}\diagup
 \sum_{\lambda _1 \lambda _2} n_{\lambda _1} (1-n_{\lambda _2})
 $$
 $$
 b=(2/3)(1-(2A)^{-1/3}),\quad
 c_A =0.8A^{-1/3}
 $$

The average energy of spin-orbit splitting $E_{ls}$ is a parameter in Eq. (2) that can be calculated either from the
single-particle scheme of GT spin-flip transitions, as was done in \cite{6,18}, or can be obtained phenomenologically by
comparing calculated $\Delta E_{\text{ G-A}}$ energies and the experimental data  \cite{16}. The dependence of $E_{ls}$ for heavy nuclei with  the
number of neutrons $N > 80$ we used the parameterization:
\begin{equation}\label{e3}
E_{ls}=20N^{-1/3}+1.25 \quad (MeV),
\end{equation}
obtained as in \cite{20} but substituting $A$ for $N$. Shell effects in the region of lighter nuclei were considered. Such
 behavior according formula (3) corresponds to decreasing the $E_{ls}$ value and to restoration of the Wigner
  SU(4)-symmetry in heavy nuclei.

 Equation (2) is also applicable for heavy and superheavy nuclei, because for them the value of  $x=\Delta E /E_{ls}$ is larger and the accuracy of the calculations should be better.

\section{Results and discussion}

Energy differences between the Gamow-Teller and analog resonances $\Delta E_{\rm G-A}=E_{\rm GTR}-E_{\rm AR}$ were calculated using
Eq. (2) for 33 nuclei: $^{48}$Ca,
${}^{60,64}$Ni, ${}^{71}$Ga, ${}^{76}$Ge, ${}^{82}$Se,
 ${}^{90,91,92,94}$Zr, $^{93}$Nb,  $^{94,96,97,98,100}$Mo, ${}^{115}$In, ${}^{112,114,116,117,118,119,120,122,124}$Sn,
 ${}^{128,130}$Te, ${}^{127}$I, ${}^{136}$Xe, ${}^{150}$Nd,
  ${}^{169}$Tm and ${}^{208}$Pb  (the initial target nuclei), for which experimental data are available (we used experimental data from \cite{16,19}).
Calculated and experimental dependences of relative energy $y(x)=\Delta E _{\rm G-A}/E_{ls}$
 on dimensionless parameter $x=\Delta E/E_{ls}$  are
presented in Fig. 1. The leftmost and rightmost points  correspond to $^{60}$Ni
and $^{208}$Pb with $x = 0.52$ and $x = 2.15$,
 respectively. The difference between the calculated and experimental results
  $\Delta \varepsilon=|\Delta E_{\rm G-A}^{calc}-
  \Delta E_{\rm G-A}^{exp}|$ is 0.38 MeV for $^{60}$Ni and less than
 0.10 MeV for $^{208}$Pb, indicating that the accuracy of calculations improves for heavy
 nuclei.
 %%%%%%%%%%%%%%%%%%%%%%%%%%%%%%%%%
 \begin{figure}
% \resizebox{1.00\columnwidth}{!}
\includegraphics{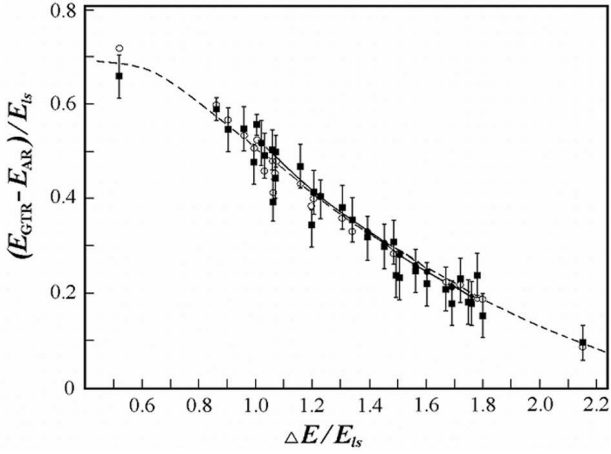}
\caption{
 Dependence of the dimensionless energy difference between the GTR and AR $y(x)=(E_{\rm GTR}-E_{\rm AR})/E_{ls}$ on parameter
$x=\Delta E/E_{ls}$ calculated by Eq. (\ref{e2}) (open circles) and experimental data (black
 squares). Black dots connected by the line represent the values calculated for Sn isotopes; the dashed line, those
  calculated with the $E_{ls}$ values obtained with Eq. (\ref{e3}) for nuclei located on the beta-stability line determent by Eq. (\ref{e4}).}
\label{fig:model_energy}
\end{figure}
The r.m.s. deviation of the calculations according to formula (\ref{e2}) for 33 listed
nuclei from the experimental data is
$\delta (\Delta \varepsilon)\leq 0.30$ MeV which is comparable to the accuracy of the $E_{\rm GTR}$ experimental data \cite{17}.

%%%%%%%%%%%%%%%%%%%%%%%%%%%%%%%%%%%%%%%%%%%%%%%%%%%%%%%%%%%%%%%%%
Fig. 1 also presents calculations for the nuclei located at the "line of
beta-stability" (LBS), which was determined by the formula
\begin{equation}\label{e4}
Z_{\beta}=A/(2+0.015A^{2/3}),
\end{equation}
 derived in \cite{21} from the condition $\partial M /\partial Z=0$ (with
 constant $A$), using well-known formula for the binding energy in the liquid-drop
  model of the nucleus. Here, $Z_{\beta}$ corresponds to the minimum mass of
  nucleus for each isobaric chain.
%%%%%%%%%%%%%%%%%%%%%%%%%%%%%%%%%%%%%%%%%%%%%%%%%
\section{Analog resonances and Coulomb displacement energies in SU(4)-approach}
%%%%%%%%%%%%%%%%%%%%%%%%%%%%%%%%%%%%%%%%%%%%%%
If the analog and Gamow-Teller resonances belong to the same supermultiplet, so the
AR energies should be described in the framework of SU(4)-theory. The analysis of
 the applicability of the SU(4)-approach was performed in \cite{10}, where the
 possibility of the description of the difference $\Delta E_{C}$ between the
  Coulomb energies of neighboring isobar nuclei within the SU(4)-theory was
  analyzed. However, this analysis was performed in [10] only for nuclei with
  $A<60$; for this reason, an unambiguous conclusion cannot be made. We analyzed
  the $\Delta E_{C}$ values for more than 400 nuclei for which experimental data
   are known in the mass number interval $A$= 3-244 (we used the data presented
   in \cite{22}). As in \cite{10,22}, we used the two parametric formula
   \begin{equation}\label{e5}
\Delta E_{\rm C}=E_{\rm C} (A,Z+1)-E_{\rm C}(A,Z)=a\frac{Z}{A^{1/3}}+b.
\end{equation}
 For all nuclei
with $A = 3-244$, we obtained $a = 1416$ and $b = -698$ keV with the standard
deviation $\delta E < 100$ keV. Deformation was taken into account phenomenologically
as in [22], by introducing the correction to $\Delta E_{C}^{def}= \Delta
E_{C}^{sph}-\delta E_{C}^{def}$ with the deformation parameters
$\beta_{2}$ and $\beta_{4}$ from \cite{23}. In the SU(4)-scheme, four types of ground
 states of nuclei belong to different supermultiplets: (i) ($Z$-even, $N$-even)
  nuclei belong to the ($T_{Z}$, 0, 0) supermultiplet; (ii) ($Z$-even, $N$-odd)
   nuclei belong to the ($T_{Z}$, 1/2, 1/2) supermultiplet; (iii) ($Z$-odd, N-even)
    nuclei belong to the ($T_{Z}$, 1/2, -1/2) supermultiplet; and (iv)
     ($Z$-odd, $N$-odd) nuclei belong to the ($T_{Z}$, 1, 0) supermultiplet where the isospin $T_{Z}=(N-Z)/2$
     and the energy $\Delta E_{\rm C}$ is considered as the difference between
     the energy of the ground state of the ($A$, $Z$) nucleus and the energy of
      excitation of the analog resonance in the ($A$, $Z + 1$) nucleus taking into
      account the energy of $\beta$-decay $Q_{\beta}$. Correspondingly,
       taking into account the mass difference $\Delta M=M_{n}-M_{\rm H}=782.35$ keV,
        we obtain the relation $b=\beta-\Delta M$ for the
        parameter $b$ from Eq. (\ref{e4}), where
        the parameter $\beta$ in the SU(4)-scheme should depend on the
         supermultiplet of the ground state. In particular, for nuclei
          with even $Z$, i.e., for cases (i) and (ii), the equality
          $\beta=0$ \cite{10} should be satisfied; this equality is
           really observed: the average deviation of the $\beta$ value from
           zero is 80 keV. The most interesting cases are $Z$-odd
            nuclei, for which the SU(4)-scheme gives the
            dependence $\beta=\alpha / T_{Z}$ on the isospin
            $T_{Z}$, where the parameter $\alpha$ is different for
             $N$-even and $N$-odd nuclei \cite{10}. The analysis shows the
             inverse dependence $\beta \approx 83/T_{Z}$ keV for nuclei with odd $Z$ values
             according to the SU(4)-approach \cite{10}. However, we were not able to obtain
             different $\alpha$ values for different supermultiplets
             because of insufficient data on $\Delta E_{\rm C}$ for
             odd-odd nuclei.

Nevertheless, analyzing the experimental data on the energies $\Delta E_{\rm C}$ for
more than 400 nuclei, one can state that the observed functional dependence
corresponds to the SU(4)-theory.

%%%%%%%%%%%%%%%%%%%%%%%%%%%%%%%%%%%%%%%%%%%%
\section{Energies of Gamow-Teller and analog resonances in heavy and superheavy nuclei}

Equations (\ref{e2}) and (\ref{e3}) are also valid for heavy and superheavy (SH) nuclei and provide even
 better results because the
parameter $x=\Delta E/E_{ls}$ is larger in this case and the conditions of the
 model approach applicability
for solving TFFS equations (\ref{e1}) are better.

%%%%%%%%%%%%%%%%%%%%%%%%%%%%%%%%%%%%55
Figure 2 shows the results of the calculations of the absolute value $\Delta E_{\rm G-A}$ as
 a function of the mass number for isotopes with $A > 140$ located on the line of beta
 stability. These isotopes with $Z_{\beta } (A)$ were found for each isobaric chain by the
 minimum mass of the nucleus from the
 experimental data \cite{24}. The microscopic calculations of energy differences between the
  Gamow-Teller and analog resonances for the $^{257}$Fm, $^{271}$Sg, $^{280}$Ds, and $^{290}$Lv isotopes with
  allowance for the single-particle structure, as in  \cite{25}, are also presented. According to
   Eq. (\ref{e1}), these calculations are approximate because the deformation of nuclei was taken
    into account phenomenologically, as in  \cite{22}. Meanwhile, the consistent inclusion of
    deformation should affect the single-particle spectrum. However, the
    effect of deformation on the energy of spin-orbit splitting, which determines the
    position of the Gamow-Teller resonance, is small. It was found that the energies
    $\Delta E_{\rm G-A}$ calculated by Eq. (1) for four heavy nuclei differ within
     0.1 MeV from those calculated by Eq. (2). As is seen in Fig. 2, the Gamow-Teller
      and analog resonances are also degenerate for heavy nuclei. However,
       the $\Delta E_{\rm G-A}$ values microscopically calculated according to
        Eq. (\ref{e1}) are somewhat larger than those obtained by Eq. (\ref{e2}) for nuclei on the
        line of beta stability because of a more correct calculation of the energy
        $E_{ls}$ from the single-particle level scheme. The analysis of spin-orbit
        splitting in superheavy nuclei performed in  \cite{26} within the generalized
        self-consistent method of the energy density functional demonstrated that the
         consistent variation of the parameters of the spin-orbit interaction
         slightly affects the energies of spin-orbit splitting; i.e., this
          quantity is stable.

Thus, taking into account degeneracy in the matrix elements of the Gamow-Teller and analog
resonances  \cite{27}, one can conclude that a decrease in the energy difference
 $\Delta E_{\rm G-A}$ between the Gamow-Teller and analog resonances in heavy
 nuclei is due to the restoration of SU(4)-symmetry and both resonances belong
  to the same Wigner supermultiplet together with the ground state of the initial ($A$, $Z$) nucleus.

\begin{figure}
% \resizebox{1.00\columnwidth}{!}
\includegraphics{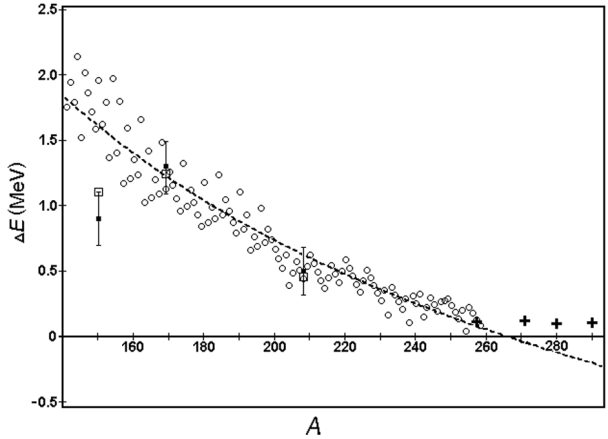}
\caption{
 Energy difference
 $\Delta E_{\rm G-A}$
 versus
 mass number $A$ (open squares) calculate
  by Eq. (\ref{e2}) and (closed squares) obtained experimentally for the $^{150}$Nd,
   $^{169}$Tm, and $^{208}$P isotopes. Circles are the calculations for nuclei
   located on the line of beta stability from \cite{24}. The dashed line is
   calculated for nuclei located on the line of beta stability determined by Eq.
   (\ref{e3}). Crosses are calculations for the $^{257}$Fm, $^{271}$Sg, $^{280}$Ds,
    and $^{290}$Lv nuclei according to Eq. (\ref{e1}).}
\label{fi}
\end{figure}

\section{Conclusions}

The calculated in this work values of the Gamow-Teller and analog resonance energy differences $\Delta E_{\rm G-A}$
 were found to be in good agreement with the experimental data. The root-mean-square deviation is 0.30 MeV for the 33 considered nuclei with known experimental data.
 The convergence of GTR and AR energies for the group of heavy nuclei with $Z \geq 100$ on the beta-stability line was investigated.
 The $\Delta E_{\rm G-A}$ values were calculated for heavy and superheavy nuclei with mass numbers up to $A=290$. On the base of observed
 degeneration of Gamow-Teller and analog resonances in heavy nuclei and predicted in superheavy,
 the Wigner SU(4)-restoration is confirmed. This allows to describe the heavy nuclei properties more confidently using SU(4) theory, especially for mass relations \cite{28}.

The developing relationship for the masses of the nuclei, and the analysis of the
Franzini-Radikatti  \cite{9} relation for nuclei masses, resulting from the SU(4)-theory,
 and which was performed for several times  \cite{29,30}, confirms that these relations works better
  in heavier nuclei. Also the analysis of the Coulomb displacement energies, using the
  SU(4)-approach allow to describe the mass and energies of superheavy nuclei with good accuracy.

So, it has been shown that Wigner supersymmetry is restored in heavy nuclei. As a result, the
 interpretation of the energy of spin-orbit splitting and the corresponding shell
 structure, as well as thereby the possibility of the existence of the "island of stability"
 in the region of superheavy nuclei, become indefinite. Our analysis of the degeneracy of
 the Gamow-Teller and analog resonances involves the ratio $x=\Delta E/E_{ls}$, which increases
  in heavy nuclei with the energy $\Delta E \sim
    (N-Z)/A$
  even at the constant value
   $E_{ls}$. Here, $E_{ls}$ is the average energy of spin-flip single-particle
   transitions within spin-orbit doublets (\ref{e2}), which decreases with an increase in the
   neutron excess. According to estimate (\ref{e2}), the energy $E_{ls}$ tends to a finite value in
    heavy nuclei and does not vanish. The microscopic calculations for superheavy
     nuclei (see Fig. 2) confirm that $E_{ls}$ is greater than zero and even increases
      slightly when approaching the "island of stability"

Thus we may conclude, that the restoration of Wigner  supersymmetry in heavy nuclei does not
contradict the possibility of the existence of the "island of stability" in the region of superheavy nuclei  \cite{31}.

\section{Acknowledgements}

The authors are grateful to S. S. Gershtein, E. E. Sapershtein, N. B. Shulgina, S. V. Tolokonnikov and D. M. Vladimirov for their assistance and helpful discussions.

This work was partly supported by the Russian Foundation for Basic Research, project no`s. 13-02-12106 ofi-m (Section 5),  14-22-03040 ofi-m (Section 4) and Swiss National Science Foundation grant no IZ73Z0\underline{ }152485 SCOPES (Section 1).

\end{document}